\documentclass[sigconf]{acmart}

\usepackage{dsfont}
\usepackage{amsmath, ulem}
\usepackage{float}
\newtheorem{theorem}{Theorem}

\theoremstyle{contribution}
\newtheorem{contribution}[theorem]{Contribution}

\AtBeginDocument{%
  }

\renewcommand\footnotetextcopyrightpermission[1]{}

\begin{document}

\title{Ranking Policy Learning via Marketplace Expected Value Estimation From Observational Data}
\titlenote{Copyright 2024 for this paper by its authors. Use permitted under Creative Commons License Attribution 4.0 International (CC BY 4.0).\\Presented at the SURE workshop held in conjunction with the 18th ACM Conference on Recommender Systems (RecSys), 2024, in Bari, Italy.}
\author{Ehsan Ebrahimzadeh, Nikhil Monga, Hang Gao, Alex Cozzi, Abraham Bagherjeiran}
\affiliation{
\institution{eBay Search Ranking and Monetization}
\country{}
}

\renewcommand{\shortauthors}{Ebrahimzadeh et al.}

\begin{abstract}
We develop a decision making framework to cast the problem of learning a ranking policy for search or recommendation engines in a two-sided e-commerce marketplace as an expected reward optimization problem using observational data. 
As a value allocation mechanism, the ranking policy allocates retrieved items to the designated slots so as to maximize the user utility from the slotted items, at any given stage of the shopping journey.
The objective of this allocation can in turn be defined with respect to the underlying probabilistic user browsing model as the expected number of interaction events on presented items matching the user intent, given the ranking context.
Through recognizing the effect of ranking as an intervention action to inform users' interactions with slotted items and the corresponding economic value of the interaction events for the marketplace, we formulate the expected reward of the marketplace as the \textit{collective} value from all presented ranking actions.
The key element in this formulation is a notion of \textit{context value distribution}, which signifies not only the attribution of value to ranking interventions within a session but also the distribution of marketplace reward across user sessions.
We build empirical estimates for the expected reward of the marketplace from observational data that account for the heterogeneity of economic value across session contexts as well as the distribution shifts in learning from observational user activity data.
The ranking policy can then be trained by optimizing the empirical expected reward estimates via standard Bayesian inference techniques.
We discuss the connections and distinctions between our proposed perspective and the standard supervised approach to learning to rank via empirical risk minimization with respect to standard information retrieval metrics.
The specific focus of this paper is to highlight the significance of the empirical context value distribution in shaping the properties of the corresponding ranking policies by contrasting various empirical importance sampling distributions.
We report empirical results from online randomized controlled experiments on a product search ranking task in a major e-commerce platform demonstrating the fundamental trade-offs governed by ranking polices trained on empirical reward estimates with respect to extreme choices of the context value distribution.
\end{abstract}

\keywords{Learning to Rank,
Counterfactual Inference,
Expected Reward Estimation,
E-commerce Marketplace}

\maketitle


\section{Introduction}
\subsection{Motivation}
Two sided e-commerce marketplaces are intermediary economic platforms that connect buyers and sellers, usually providing a wide selection of products for the buyers from a diverse array of sellers.
The primary \textit{buyer focused} objective of the marketplace is to guide buyers through their browse and discovery journeys to identify and purchase the items that fulfill their \textit{shopping intent}.
Users’ browsing and purchase journeys in the marketplace are impacted by an ecosystem of decision making systems, most notably via the ranking policies in various stages of their shopping journeys from discovery pages to the Search Engine Result Pages(SERP).
An effective ranking policy aims to showcase a set of results matching the user intent at any given ranking context along the shopping journey with rewards realized as interaction events on the slotted items on the page.
Collectively, user journeys are not equally likely to produce value for the marketplace and the goal is to expand the set of successful user sessions, optimizing a suitable notion of long term value for the users and the marketplace.
It is therefore essential for the ranking policy to account for the utility of all stakeholders in this economic setting.
In standard formulations of learning to rank in the information retrieval literature, however, there is usually no clear connection between the training objective for the ranking policy, the long term value for the collective of the users and the key performance metrics of the marketplace.  
In this paper, focusing primarily on the search ranking policy invoked in response to users' search queries, we formulate the ranking policy learning as an optimization problem based on a (counterfactual) estimate of the marketplace reward from observational data.
\subsection{Contributions and Related Work}
\begin{contribution}
We propose a decision making framework establishing explicit connections between learning a ranking policy for a search/recommendation engines and building effective empirical estimates for a suitable notion of marketplace expected reward.  
\end{contribution}
The problem of developing merit scores for ranking items, post a selection stage from a large pool of candidates, is widely studies in the context of recommendation systems\cite{ma2018entire}, display advertising\cite{chaudhuri2017ranking}, sponsored search\cite{zhang2014sequential}, and search ranking\cite{wu2018turning}, where the sequential and hierarchical nature of the user interaction events and sparsity of success events\cite{wang2022escm2,ma2018entire,jin2022multi,o2021analysis} in user journeys are taken into account.
Value-aware policies in the context of advertising\cite{zhang2014sequential,zhang2014sequential}, and economic recommender systems\cite{de2023economic,de2023systematic,ma2018entire} account for business objectives, primarily through manipulations of the merit scores based on conversion likelihood estimates and the price of the candidate items to develop a point-wise notion of expected value for a given candidate item. 
In contrast, our approach is user focused in that the goal of the ranking policy is to optimize for the user utility in the sense of maximizing the expected number of engagements on desirable items at every stage of the search journey.
An alternative formulation is to frame the search ranking policy learning as a multi-stakeholder multi-objective optimization problem\cite{mahapatra2023multi,carmel2020multi}, with potentially conflicting objectives\cite{tsagkias2021challenges} that account either for business constraints\cite{tang2024multi} or group exposure constraints\cite{mehrotra2018towards}.
The notion of value for the marketplace is introduced in the ranking policy objective via an importance weighting distribution that signifies both the economic value and the likelihood of realizing some reward from an interaction event with an item that satisfies the user intent. 
\begin{contribution}
We characterize the key elements in building effective (policy-dependent) expected reward estimates from observational data, controlling for (1) the heterogeneity of the session value distribution, (2) the contribution of interventions within a user journey via the reward attribution scheme, and (3) the distribution shifts incurred by selection biases in observational data.
\end{contribution}

Reinforcement learning(RL) is a powerful framework to account for sequential interventions within the session by formulating the problem of recommending new items\cite{shani2005mdp} or search ranking\cite{hu2018reinforcement}  as a Markov Decision Process (MDP). 
By expanding the planning horizon and adopting intermediary reward shaping techniques, RL-based approaches account for delayed rewards in the session, via suitable representations of the dynamic session context(state) in session trajectories.  
Recognizing the selection biases in the observed user behavior data, offline reinforcement learning techniques, including inverse propensity weighting\cite{chen2019top}, and actor-critic methods\cite{chen2022off}, are adopted to account for distribution shifts in learning from logged data. Similar counterfactual training techniques based on propensity weighting and potential outcome modeling are developed in the context of counterfactual learning to rank for search ranking problems\cite{joachims2017unbiased,ebrahimzadeh2023counterfactual,oosterhuis2023doubly}.
There is, however, no clear account of the heterogeneity of marketplace reward across session trajectories, neither in the standard counterfactual supervised learning perspective nor in offline reinforcement learning approaches.
\begin{contribution}
We highlight the significance of the empirical session-context value distribution in building effective marketplace expected reward estimates by demonstrating fundamental performance trade-offs governed by the search ranking policies trained on extreme choices of the context value distribution via rigorous counterfactual evaluations as well as online randomized controlled experiments in a major e-commerce platform.     
\end{contribution}
The definition of success events and the associated reward to the user events is flexible in our framework and is informed by the strategic choices of the marketplace.
Specifically, an early-stage marketplace may focus on maximizing the collective number of engagements, while an acquisition-oriented marketplace targets the collective number of purchases, while a revenue-driven marketplace chooses to maximize the long-term gross merchandise value.

\subsection{Notation}
Here is a list of notation adopted throughout the paper.
Sets and ordered sets(lists) are represented with upper-case calligraphic symbols; such as $\mathcal{X}$. 
Random quantities are shown in bold such as $\mathbf{x}$ with realization $x$. 
The expected value of random variable $\mathbf{x}$ is denoted by $\mathbb{E}[\mathbf{x}]$
and the conditional expectation of a random variable $\mathbf{z} = f(\mathbf{x},\mathbf{y})$ given $\mathbf{y}$ is denoted by $\mathbb{E}[\mathbf{z}|\mathbf{y}]$ or $\mathbb{E}_{\mathbf{x}\sim\mathbb{P}(x)}[\mathbf{z}]$.
For a function $f:\mathbb{X} \to \mathbb{R}$, the $|\mathcal{X}|$ dimensional array $[f(x)]_{x \in \mathcal{X}}$ is denoted by $f(\mathcal{X})$.

\section{Problem Setup}
\subsection{Decision Making Framework}
The marketplace is interested in maximizing the average total reward across all user session trajectories over a long time horizon
\begin{equation}
\frac{1}{T}\sum_{t\leq T}v_{s_t},
\end{equation}
where $v_{s}$ is the economic value from a successfully served search session $s$.
Our framework is flexible in the choice of the reward function and we discuss the fundamental trade-offs between multiple strategic marketplace long term reward choices, namely revenue-based, value per engagement and value per acquisition marketplaces.  
The reward from a session trajectory is assumed to be non-negative. Although our framework, can be extended to account for negative rewards, we ignore it in our formalization.
We assume that the reward over search journeys is a stationary ergodic stochastic process.
By invoking Birkhoff’s ergodic theorem\cite{durrett2019probability}, with probability $1$, the long term temporal average is same as the expected reward, i.e. 
\begin{equation}
\mathbb{E}_{\mathbf{s},\mathbf{v}} [\mathbf{v}_\mathbf{s}],  
\end{equation}
where the expectation is taken with respect to the randomness in the session context and reward distribution.

While our formalization can naturally be extended to search journeys with complex goals, we focus on a typical e-commerce purchase decision making scenario of session trajectory a with a single product intent, ignoring sessions with multi-product purchase intent, as well as informational and navigational search sessions.
We recognize that users' decisions are impacted by multiple independently optimized decision making systems, but we are oblivious to potential interactions of the ranking policy with these systems, specifically to the closely related query understanding and candidate retrieval policies.
We only focus on policy learning for search result pages with a single layer presentation semantic where the action of the ranking policy is the permutation/ranking of a largely homogeneous set of \textit{comparable items} for a flat single-layered presentation of the results, ignoring the multiplicity of user's search intent and diversity considerations for the result set.

We cast the problem into a Bayesian decision making framework with a user-focused perspective on the definition of success upon a ranking action.
The ranking policy aims to increase the expected number of engagements on items that meet the user intent, and the reward is proportional to the likelihood of a success event(non-zero reward) from the user interactions on the search results page produced by the ranking policy.
A crucial aspect of this framework is to account for distribution shifts in observational data, i.e. the distinction between the distribution of the logged search activity data that the policy is trained and the inference time distribution of user events.

\subsection{Success From a Ranking Intervention}
\label{section_success_ranking_intervention}
Given a search query $q$ within a session context $s$, the ranking policy $\pi:\mathcal{D}_q \to \{1,\cdots,N\}$  maps a candidate item d from the retrieved set $\mathcal{D}_q$ to a slot $\pi(d)$. 
The notion of success with respect to a slotting $\pi(\mathcal{D}_q)$ of the items on the SERP $q$ is defined based on the effectiveness of the policy in driving user interaction events(Click).
Specifically, the objective of the ranking policy on a given ranked SERP is to increase the expected number of engagements on desirable items(suitably defined) $c_{\pi(\mathcal{D}_q)}$ given the session context upon issuing the query $s_{\prec q}$; i.e.
\begin{equation}
    \mathbb{E}[\mathbf{c}_{\pi(\mathcal{D}_q)}|s_{\prec q}],
\end{equation}
where the expectation is with respect to the randomness in user preferences and browsing behaviors in the given query context and possibly the randomness in the ranking policy(if stochastic). Note that $s_{\prec q}$ subsumes all the relevant contextual information upon issuing the query $q$; including all the queries and the corresponding surfaced items, as well as the engaged items prior to the current query context.  
In the subsequent sections, we will make probabilistic assumptions on users' browsing and click behaviors and build effective policy dependent empirical estimators.

\subsection{Success From a User Session}
\label{section_success_user_session}
The notion of success with respect to a user session is defined based on interaction events on desirable items across all interventions by the marketplace within a user journey. 
Given the per query ranking objective $\mathbb{E}[\mathbf{c}_{\pi(\mathcal{D}_q)}|s_{\prec q}]$ (i.e. the expected number of desirable engagements from the SERP given the session context upon issuing the query), the success from the overall user session is shaped by the distribution $\mathbb{P}(q|s)$ that signifies the contribution of the interactions on the ranked SERP $q$ to the overall success of the user search session $s$. 
\begin{equation}
    \mathbb{E}_{\mathbf{q}\sim \mathbb{P}(q|s_{\prec q})}[\mathbb{E}[\mathbf{c}_{\pi(\mathcal{D}_{\mathbf{q}})}|s_{\prec \mathbf{q}}]],
\end{equation}
This distribution is usually referred to as the \textit{success attribution distribution}, which is primarily studied in the context of online advertising\cite{ji2016probabilistic}.
The key difference is that in online advertising the unit of value attribution is item impression, while in this work we emphasize the attribution of value to the \textit{ranked list} shown for the query given the prior context. 
This is also related to the \textit{credit assignment problem} in reinforcement learning on how to attribute success to the intermediate actions of the agent.

\subsection{Marketplace Expected Reward}
\label{section_marketplace_expected_reward}
The expected reward of the marketplace from the presented ranking is then shaped by the distribution of the value across search contexts, which signifies the economic value of the user-interaction events in the sessions for the marketplace. 
The random variable $\mathbf{v}_\mathbf{s}$ captures the strategic notion of the value of the session for the marketplace, which corresponds to the value of the interactions events on the item(s) that satisfy the user's intent.
The expected reward of the marketplace can then be written as
\begin{equation}
   \mathbb{E}_{\mathbf{s},\mathbf{v}}[\mathbf{v}_\mathbf{s}\mathbb{E}_{\mathbf{q}\sim \mathbb{P}(q|s_{\prec q})}[\mathbb{E}[\mathbf{c}_{\pi(\mathcal{D}_{\mathbf{q}})}|\mathbf{s}_{\prec \mathbf{q}}]]],
\end{equation}
We can also consider an alternative formulation where we assume that session value distribution $\mathbb{P}_v(s)$ subsumes both the \textit{likelihood of the user session to lead to some reward} for the search engine as well as the \textit{reward value} attributed to the user session $s$:
\begin{equation}
\label{populationExpectedReward2}
       \mathbb{E}_{\mathbf{s}\sim \mathbb{P}_v(s)}[\mathbb{E}_{\mathbf{q}\sim \mathbb{P}(q|s_{\prec q})}[\mathbb{E}[\mathbf{c}_{\pi(\mathcal{D}_{\mathbf{q}})}|\mathbf{s}_{\prec \mathbf{q}}]]].
\end{equation}
For a value-aware search engine with marketplace revenue objective, economic value is realized only in the event of a transaction as the success event from a search session and the reward is proportional to the price of the sold item(s).
For a search engine that aims to optimize for the volume of transactions, it is more suitable to adopt a value per acquisition notion of reward oblivious to the price of the sold items. 
For a search engine with strategic goal of maximizing user engagements for increased user retention and minimizing abandonment, it is more suitable to adopt a value per click notion of reward oblivious to the post click transaction events.

In the next section, we discuss empirical modeling techniques to build effective empirical reward estimates from observational data, which effectively frame the problem as a standard counterfactual empirical risk minimization with a value-aware context distribution. 
\section{Expected Reward Estimation from Observational Data}
\label{section3}

\subsection{Estimating the Per Query Success}
\label{section_estimate_per_query_success}
We are interested in maximizing $\mathbb{E}[\mathbf{c}_{\pi(\mathcal{D}_q)}|s_{\prec q}]$, the expected number of \textit{desirable engagements} across all the slots on the page, with a suitably parameterized ranking policy $\pi$.
By hypothesizing an explanatory click model based on causal constructs that govern user browsing and engagement behaviors on search result pages, we can build effective likelihood models from which we can estimate the parameters of the ranking policy via maximum likelihood estimation using logged observational data. 
We instantiate this process with the simple widely adopted click models in information retrieval.
Assuming a vanilla Sequential Browsing Model along with the standard Position-Dependent Examination Model, we can write the expected number of desirable engagements as
\begin{eqnarray}
\mathbb{E}[\mathbf{c}_{\pi(\mathcal{D}_q)}|s_{\prec q}]\!\! &=& \!\!\!\sum_{r=1}^N \mathbb{P}(\mathbf{c}_{\pi^{-1}(r)}=1|s_{\prec q}, \mathcal{D}_{q}^{\prec r})\\
\!\!&=& \!\!\!\sum_{r=1}^N \mathbb{P}(\mathbf{c}_{\pi^{-1}(r)}=1|s_{\prec q},r)\\
\!\!&=& \!\!\!\sum_{r=1}^N \mathbb{P}(\mathbf{o}_{\pi^{-1}(r)}=1|q)\!\times\! \mathbb{P}(\mathbf{R}_{\pi^{-1}(r)}=1|s_{\prec q})\\
\!\!&=& \!\!\!\sum_{r=1}^N \mathbb{P}(\mathbf{o}_{\pi^{-1}(r)}=1)\!\times\! \mathbb{P}(\mathbf{R}_{\pi^{-1}(r)}=1|s_{\prec q})
\end{eqnarray}
where the first line follows from sequential browsing assumption with $\mathbf{c}_{\pi^{-1}(r)}=1$ representing the click event on the item ranked at position $r$ and $\mathcal{D}_{q}^{\prec r}$ representing the slotted items prior to the item ranked in position $r$. 
The second line follows from assuming that the user interaction event on a given slot is independent of the placed items in the previous slots.
The third line follows from the standard examination-based click model that posits that a click event can be expressed as the intersection of a query specific examination event $\mathbf{o}_{\pi^{-1}(r)}=1$ and a presentation-independent contextual relevance event $\mathbf{R}_{\pi^{-1}(r)}=1$; and the last line follows from assuming a query-independent global rank discount function on the examination probabilities.
The examination probabilities, a.k.a. propensity scores, are context-specific and can be estimated via explicit online interventions or from observational data.
By considering a simple uni-variate model fit on the estimated propensities as a function of rank, one can build data-driven rank-discount functions to estimate users' examination effort.
However, the standard approach is to adopt vanilla log-based context-oblivious rank discounts as generic estimates of the examination probability  $\hat{\mathbb{P}}(\mathbf{o}_{\pi^{-1}(r)}=1)$, i.e.
$$\ell(r)=\frac{1}{\log(1+r)}.$$ 
Given an empirical estimate $\hat{r}$ of Bayes contextual relevance probabilities $\mathbb{P}(\mathbf{R}_{\pi^{-1}(r)}=1|s_{\prec q})$, one can derive the standard discounted cumulative gain (DCG) estimate $\hat{\mathbb{E}}_{\mathrm{DCG}}[\mathbf{c}_{\pi(\mathcal{D}_q)}|s_{\prec q}]$ of the expected reward per query for the policy $\pi(\cdot)$ as
\begin{eqnarray}
\label{DCG_formula}
 \ell(\pi(\mathcal{D}_q))^T  \hat{r}(\mathcal{D}_q).
\end{eqnarray}

Upon building all the elements of the expected reward estimates, specifically the policy-dependent per query expected reward estimates, we can train the ranking policy by maximizing this empirical reward estimate, as elaborated in the next section.  

\subsection{Estimating the Session Expected Reward}
\label{section_estimate__expected_reward}
\subsubsection{In-session success attribution}
Several techniques can be adopted to estimate the contribution of a ranked SERP $q$ and the corresponding \textit{observed or potential} interactions on that page to the overall success of the user search session $s$.
A simple yet popular solution in the context of online advertising is to adopt an attribution distribution that assigns all the probability mass to the immediate query context preceding the post-click conversion event, which is referred to as Last Touch Attribution scheme.
In contrast to this tight attribution scheme, one can assume a uniform distribution across all queries in the session in which the item with the attributed interaction event of interest was retrieved as a candidate item, oblivious to whether it was even impressed on the search result page. 
This approach is referred to as All Touch attribution scheme.
Alternatively, One can assume a (Markovian) probabilistic graphical model on user's touch points within a session journey and infer a probabilistic multi-touch attribution distribution 
$$\hat{\mathbb{P}}(q|s_{\prec q})$$
from observational data. 
Similarly, one can adopt an attention based sequence modeling approach and infer the contribution weights for interaction events along the user journey with a conversion prediction model.   
Lose attribution schemes, like the all touch attribution scheme, signify the powerful idea of \textit{counterfactual training context generation} for ranking policy learning, where in contrast to predictive perspectives, the policy can collect reward from a ranking context where the item of interest was not observed by the user. 
As discussed in the empirical results section, such attribution schemes are particularly effective for capturing the user behavior in search sessions with longer feedback loops, e.g. sessions with high purchase value user intent.    

\subsubsection{Session Value estimation}
In order to highlight the importance of the session value distribution, 
$$\hat{\mathbb{P}}(s)$$
let us focus on a search engine with a value per acquisition objective.
A straightforward empirical session value distribution is adopt a uniform distribution on sessions that lead to a transaction event.
Such session value selection distribution leads to survivorship bias in training context selection in that traffic segments where transaction events are rare, e.g. user sessions with luxury intent, will be under-represented in training.
A simple approach is to expand the definition of success events and estimate the likelihood of session success with a content-oblivious estimate based on the aggregate conversion likelihood of the \textit{richest engagement event} attributed to the element(s) engaged.
This perspective on building mixture distributions based on the richest post-click engagement event was shown to be effective in capturing potential conversions from browse-heavy user journeys\cite{seyler2023aligning}.

For a revenue-focused marketplace, as discussed in Section \ref{section_success_ranking_intervention}, the value of a search session is proportional to the price of the item that matches the user intent. In the presence of an observed success event in the logged data, the purchase price of the item to which the success event is attributed is the realization of the session value; otherwise, in the absence of an interaction event, the value of the session has to be estimated from the content of the asked intent in user queries, or a Canonical set of actual or synthesized items that match the user intent.
\subsection{Selection Bias Correction}
One of the main challenges in learning from observational data is the distribution shift between the training data collected from the logging policies and the inference data distribution.
We therefore have to introduce another set of techniques, e.g. importance weighting distributions, to account for this mismatch between the (population) expected reward
in (\ref{populationExpectedReward2}) and the estimated expected reward from the estimated quantities in the previous sections; that is,
\begin{equation}
       \mathbb{E}_{\mathbf{s}\sim \hat{\mathbb{P}}(s)}[\mathbb{E}_{\mathbf{q}\sim \hat{\mathbb{P}}(q|s_{\prec q})}[\hat{\mathbb{E}}[\mathbf{c}_{\pi(\mathcal{D}_{\mathbf{\mathbf{q}}})}|\mathbf{s}_{\prec \mathbf{q}}]]].
\end{equation}
An important source of distribution shift in observational search activity data is the the selection bias due to presentation of the items on the page and the sequential browsing of the users, implying that we only observe relevancy of the items to the user only in the event of an explicit user engagement and it is more likely to observe engagements on SERPs from higher ranking slots. 

A key technique to account for this effect is to define a suitable notion of propensity, which is developed in the context of studying the effect of a treatment(an intervention) on a population by taking into account attributes of the treatment unit in the way the treatment is assigned. In the context of ranking, the treatment is defined in correspondence to the examination of a slotted item by the user, but the key difference with the standard applications of this concept is that the examination variable is not fully observable.
An alternative approach based on potential outcome modeling, similar to actor-critic networks in the context of offline reinforcement learning, is proposed in \cite{ebrahimzadeh2023counterfactual}, where distilled knowledge from a teacher model is used in the form of soft predicted relevance labels to account for unobserved user feedback to achieve variance reduction and improved generalization. 

\subsection{Variance Reduction and Generalization}
Having discussed an array of importance weighting schemes to build empirical expected reward estimates, it is essential to develop variance reduction techniques to control the generalization behavior of expected reward estimators.
For brevity of presentation, we briefly discuss the various reduction techniques adopted and ignore developing generalization bounds on the bias and variance of the estimation error of the proposed empirical reward estimation techniques.

\subsubsection{Truncation and Bucketing}
\label{TruncationBucketing}
Clipping and truncated importance sampling techniques\cite{bottou2013counterfactual,ionides2008truncated} are popular techniques to control the variance and generalization behavior of inverse propensity weighting estimators when there is high variance in the estimated propensities. Since we combine multiple importance sampling techniques to account for selection bias, success likelihood, and context value distribution across highly heterogeneous user trajectories, we adopt this simple variance reduction technique off the shelf.

In building empirical session value distributions for a revenue focused marketplace reward, relying on the purchase price of the success items leads to a very high variance estimator, particularly in the presence of high heterogeneity in price intent across user trajectories.
Instead, we can use a stratification technique by bucketing user sessions based on value buckets defined according to the empirical revenue distribution.
Specifically, we can build a session value distribution based on the empirical revenue share of the bucket corresponding to the price of the purchased item.

\subsubsection{Potential Outcome Modeling}
One of the the primary challenges of counterfactual learning to rank from logged search activity data is that the relevancy of the items is observed only in the event of explicit user engagements. 
A popular idea in the context of contextual bandits and recommendation systems to circumvent the challenges in this partial information setting is to use predictive models for reward estimates as potential outcome models in conjunction with inverse propensity weighting\cite{wang2018deconfounded,dudik2011doubly,schnabel2016recommendations}.
There are a number of recent works in the context of unbiased response prediction that leverage and analyze the doubly robust technique\cite{zou2022approximated,wang2019doubly,saito2020doubly,oosterhuis2023doubly}.
In \cite{ebrahimzadeh2023counterfactual}, a generalized form of potential outcome modeling is proposed where the distilled knowledge from a relevance teacher is used in the form of soft predicted relevance labels to help the student with more effective list-wise comparisons, variance reduction, and improved generalization behavior.
This is similar to the idea of actor-critic networks in the context of offline reinforcement learning\cite{chen2022off}, and augmentation policy in the context of contextual bandits\cite{tucker2023variance}.
Using knowledge distillation helps build training contexts from logged search contexts without user interaction events leveraging complex models. To simplify the discussions, we ignore discussing any details about the teacher models used in our experimental setup. 

\subsubsection{Stratification and Normalization}
Effective stratification is a key technique in the context of importance weighting estimators, e.g. the context value binning idea discussed in sub-section \ref{TruncationBucketing} or training context stratification based on characteristics of logged training contexts\cite{ebrahimzadeh2023intent}.
We adopt Self-Normalizing propensity based estimators, recently analyzed in \cite{london2023self}, where we use engagement ranks as yet another stratification dimension in our proposed estimators.
Yet another standard variance reduction technique that we adopt to control the contribution of the search sessions with many success events in the observational data is to adopt normalization techniques; e.g. the standard Ideal cumulative gain normalization for the per query loss. We note that under this cumulative reward normalization technique, per item propensity weights should be reformulated as context weights.

Having equipped our empirical reward estimates with variance reduction techniques, from this point on, we can assume that the effect of all importance weighting schemes discussed so far are reflected in importance weights $\hat{v}_{q,s}$.

\subsection{Optimization Objective for the Ranking Policy}
We consider deterministic policies parameterized by a scoring function $f$, such that $\pi_f = \mathrm{argSort(f)}$, oblivious to the representation of the items and the ranking context. 
An appealing approach, particularly in the context of online advertising and sponsored search, is to directly estimate the Bayes contextual relevance probabilities $\mathbb{P}(\mathbf{R}_{d}=1|s_{\prec q})$, or equivalently the counterfactual probability of click had the item been examined $\mathbb{P}(\mathbf{c}_{d}=1|s_{\prec q},\mathrm{do}(\mathbf{o}_d=1))$ via a standard supervised predictive models, i.e.
\begin{eqnarray}
\sum_{s,q} \hat{v}_{q,s} \sum_{d\in D_q} \mathrm{D}(f(d)|| \hat{r}_d),
\end{eqnarray}
where $\hat{v}_{q,s}$ is the empirical importance weights based on discussion in section \ref{section_estimate__expected_reward}, $D$ is a distance measure, e.g. cross entropy, between the predicted distributions $f(d)$ and the properly debiased empirical label distribution$\hat{r_d}$.
For estimating counterfactual probability of click that is \textit{Contextually Well-Calibrated and Discriminative} for ranking, we need very complex models with rich feature representations, with careful data stratification and selection bias correction. Since absolute merit estimation is usually a harder problem than difference in merit estimation, we resort to alternative techniques for empirical expected reward optimization.

The standard alternative approach is to adopt the LambdaLoss framework\cite{wang2018lambdaloss} and optimize a pairwise upper bound on the (list-wise) empirical estimates for the expected number of engagements, $\ell_q(\pi_f,\hat{r})$, to circumvent the challenges of dealing with highly non-smooth rank-dependent policy function, which can be written as 
\begin{eqnarray}
   \ell_q(\pi_f,\hat{r}) = \sum_{d,d' \in \mathcal{D}_q} \Delta \hat{\mathbb{E}}_{\pi_f}(\mathrm{swap}_{\hat{r}}(d,d')) \sigma(f(d)-f(d')),
\end{eqnarray}
where $\hat{\mathbb{E}}_{\pi_f}(\mathrm{swap}_{\hat{r}}(d,d'))$ is the difference in the estimated expected number of engagements had the ranked slots of the item pairs $(d,d')$ been swapped and $\sigma(\cdot)$ is some inverse link function, e.g. softMax. 
The approximate surrogate objective, suitably weighted with the empirical reward estimates $\hat{v}_{q,s}$, expressed as
\begin{eqnarray}\label{valueweightedreward}
   \sum_{s,q} \hat{v}_{q,s} \ell_q(\pi_f,\hat{r}) 
\end{eqnarray}
can then be optimized using iterative optimization techniques, like Expectation-maximization; that is given an estimate $f^{(t)}$ at iteration $t$, in order to build $f^{(t+1)}$ from the gradient updates from the objective function, the difference in estimated objective $\hat{\mathbb{E}}_{\pi_{f^{(t)}}}$ from the swap operation is computed based on ranking order produced by $f^{(t)}$. 

\section{Evaluations and Discussions}
In Section \ref{section3}, we discussed essential elements of building empirical expected reward estimates for training effective search ranking policies.
Since conducting thorough ablation studies for characterizing the effect of each element in building empirical expected reward estimates is not possible given the space constraints, we focus primarily on the rather under-explored element in the literature, which is the effect of \textit{context value distribution} discussed in section \ref{section_estimate__expected_reward} in shaping the properties and the generalization performance of the ranking policy.

We focus on a product search ranking scenario in a major E-commerce platform and evaluate candidate policies via online randomized control experiments, as well as rigorous counterfactual evaluations on user session data collected from the online traffic.
Since all experiment are performed on proprietary data, we only report lifts compared to a simple clearly-specified baseline, with a focus on the relevant choices for controlling the \textit{estimation error} with respect to the research question of interest, oblivious to the optimization framework, the feature representations, and the hypothesis class. Specifically, we only discuss the choice of the ranking objective and the relevant importance sampling and attribution techniques for building our estimators of interest, without discussing the details of the models.

\subsection{Online Evaluation Framework}
Since the main goal of the proposed decision making framework is to build search ranking policies that generalize with respect to a given notion of marketplace expected reward, we primarily evaluate the performance of the candidate policies in online randomized controlled experiments. 
Specifically, we adopt an experiment design and primary success metric defined with respect to lifts in cumulative reward in treated user sessions.
This cumulative reward driven design is in contrast to the standard experiment design practices for incremental ranking changes, where the primary success metric is set to be the standard (immediate) ranking efficiency metrics that measure concentration of success events in Top slots, through simple attribution and aggregation schemes across search result pages.
In fact, top slot engagement concentration metrics, e.g. per query DCG with respect to SERP interactions aggregated uniformly across all queries, which are usually tightly correlated with the marketplace reward, should only be treated as secondary metrics in the presence of a measurement of cumulative reward in online experiments.
We do recognize, however, that DCG-type metrics are particularly crucial for counterfactual off-policy evaluations, as approximations to the per query expected reward using logged data, because all we can do is to measure concentration of logged success events in top slots upon the shuffling action of the new target policy.

We establish the fundamental trade-offs between ranking policies trained on different empirical expected reward objectives primary based on session level cumulative reward metrics, including Number of Engagements, Number of Purchases, and Revenue, as measured in online AB tests. For metrics that attribute the observed effect to search events, we use a simple attribution schemes based on the immediate Search Result Page that precedes the user event of interest.
\subsection{Training Objectives and Offline Evaluation Metrics}
We adopt the standard supervised counterfactual training and evaluation framework based on logged search activity data collected from the online traffic of a major E-commerce platform.
We are oblivious to the logging policy and collect datasets with importance sampling and reward attribution semantics based on the corresponding notions of expected reward of interest.
Specifically, given a target notion of expected reward, the context value distribution remains the same for training and evaluation datasets.
For candidate item selection per SERP, however, we sample three negative samples at random from impressed unengaged items within each training context, but keep all the candidate items to be re-ranked by the candidate ranker for the evaluation datasets.

For all empirical expected reward metrics, we use the same, suitably debiased and normalized, DCG approximation for the per query expected reward according to (\ref{DCG_formula}).
Unless explicitly stated otherwise, we use the following vanilla empirical context value distribution for building expected reward estimates as training objectives and the counterfactual metrics.

\textbf{Expected number of engagements $\hat{\mathbb{E}}[\mathrm{C}]$}: The session value distribution $\hat{\mathbb{P}}_{\mathcal{C}}(s)$ is a uniform distribution across logged sessions with at least one click event. We consider a simple last touch attribution scheme $\hat{\mathbb{P}}_{\mathcal{C}}(q|s_{\prec q})$ for the distribution of reward among queries within the session. 

\textbf{Expected number of purchases $\hat{\mathbb{E}}[\mathrm{P}]$}: The session value distribution $\hat{\mathbb{P}}_{\mathcal{P}}(s)$ is uniform across logged sessions with at least one purchase event. We use a simple multi-touch attribution scheme $\hat{\mathbb{P}}_{\mathcal{P}}(q|s_{\prec q})$ with uniform distribution across all queries in the converting session, where the purchased item appeared as a candidate.

\textbf{Expected revenue $\hat{\mathbb{E}}[\mathrm{Rev}]$}: The session value distribution $\hat{\mathbb{P}}_{\mathcal{R}}(s)$ is defined on the sessions with a transaction event according to the empirical revenue share of the bucket corresponding to the price of the purchased item. The same multi-touch attribution from above is adopted for this reward estimate as well.

To best highlight the heterogeneity of user behavior with respect to the underlying shopping intent and the associated fundamental trade-offs between different notions of marketplace expected reward, we also stratify our evaluations across traffic segments defined based on purchase price intent of the users, as realized in the price of the purchased item.
The price intent bins are defined in such a way so that the empirical revenue distribution is roughly uniform across value buckets.

\subsection{Research Questions}
\subsubsection{Marketplace Reward Trade offs}
The primary insight that we would like to highlight in our evaluations is the heterogeneity of users' browsing and shopping intents, as reflected in different notions of marketplace reward from user sessions.
These observations signify the crucial importance of the choice of the empirical session value distribution in shaping the properties and the generalization behavior of the search ranking policy.

We do this by contrasting the performance of ranking policies trained on expected reward estimates corresponding to extreme choices of the empirical session value distribution.
Specifically, we compare a policy $\pi_{\mathcal{C}}$, corresponding to a scoring function $f_{\mathcal{C}}$, trained on a simple engagement-driven expected reward estimate based on the session value distribution $\hat{\mathbb{P}}_{\mathcal{C}}(s)$ against a policy $\pi_{\mathcal{P}}$, corresponding to a scoring function $f_{\mathcal{P}}$, trained on a simple acquisition-focused expected reward estimate based on the session value distribution $\hat{\mathbb{P}}_{\mathcal{P}}(s)$.
We observe meaningfully different performance trade-offs between these extreme policies with respect to the primary notions of marketplace reward in an online randomized controlled experiment.
Table \ref{tab:AB_click_sale} summarizes the key observations on the average effect size $\Delta_{\mathcal{S}}(m_{\pi_{\mathcal{C}}},m_{\pi_{\mathcal{P}}})$ between engagement focused policy $\pi_{\mathcal{C}}$ and acquisition focused policy $\pi_{\mathcal{P}}$, with respect to different cumulative metrics $m$, over the global session traffic $\mathcal{S}$.
 The main takeaway from these observations is that the engagement focused policy $\pi_{\mathcal{C}}$, on the one hand, drives significantly higher share of search sessions with at least one click($>+3\%$), and on the other hand, leads to a significant drop in the share of search sessions with at least one purchase($<-2\%$).\vspace{-0.25cm}
\begin{table}[H]
    \centering
    \begin{tabular}{| c | c |}
    \hline
     Online Metric $m$  &  $\Delta_{\mathcal{S}}(m_{\pi_{\mathcal{C}}},m_{\pi_{\mathcal{P}}})$ \\ \hline
      Sessions With Any Engagement &  >+3\% \\ \hline
      SERPs With Any Engagement  &  >+5\%\\ \hline
      Sessions With A Purchase Event  &  <-2\% \\ \hline
      Total Revenue & neutral \\ \hline 
    \end{tabular}
    \caption{\small Online AB test results contrasting policies with extreme session value distributions}
    \label{tab:AB_click_sale}
\end{table}
\vspace{-0.85cm}It is interesting to note, however, that this drop is largely due to a significant loss in the number of bought items in search sessions with lower price intent, which usually take less exploration and browsing to identify and pinpoint the desirable item to purchase.
Since the engagement-driven policy is more effective in driving success events with higher economic value in sessions that require more browsing effort, it can compensate for the revenue loss due to lower purchases in lower price intent segments, leading to an overall neutral effect size in total revenue.

In order to explore the fundamental trade-offs, highlighted in our online experiment, between different notions of marketplace expected reward across heterogeneous price intents in more depth, we build simple hybrid policies corresponding to a mixture of the engagement-based and acquisition based objectives.
Specifically, we build a simple policy via a simple convex combination of the extreme polices
\begin{eqnarray}
    \pi_{\alpha} = \mathrm{argSort}((1-\alpha) \,f_{\mathcal{P}}+ \alpha \,f_{\mathcal{C}}),
\end{eqnarray}
where $\pi_{\alpha}$ refers to the balanced ranking policy obtained via a linear combination of the scoring functions of the engagement focused policy, $\pi_{\mathcal{C}}$ and acquisition focused policy $\pi_{\mathcal{P}}$, for some $\alpha \in [0,1]$. 
The parameterized policies $\pi_{\alpha}$ behave similarly to a policy trained on a corresponding mixture session value distribution $(1-\alpha)  \hat{\mathbb{P}}_{\mathcal{P}}(s)+ \alpha \hat{\mathbb{P}}_{\mathcal{C}}(s)$.

Due to the scarcity of online experimentation traffic, we only conduct counterfactual off-policy evaluations for these parameterized policies.
While our counterfactual estimates are largely aligned, at least directionally, with the measured effect sizes in online experiments, we point out that all counterfactual off-policy evaluations are fundamentally limited having access only to snapshots of the users' behavior in the logged sessions.
In particular, if the logging policy is substantially different from the target policy to be evaluated, the offline evaluation metrics could be very biased.\vspace{-0.25cm}
\begin{figure}[H]
    \includegraphics[scale=0.18]{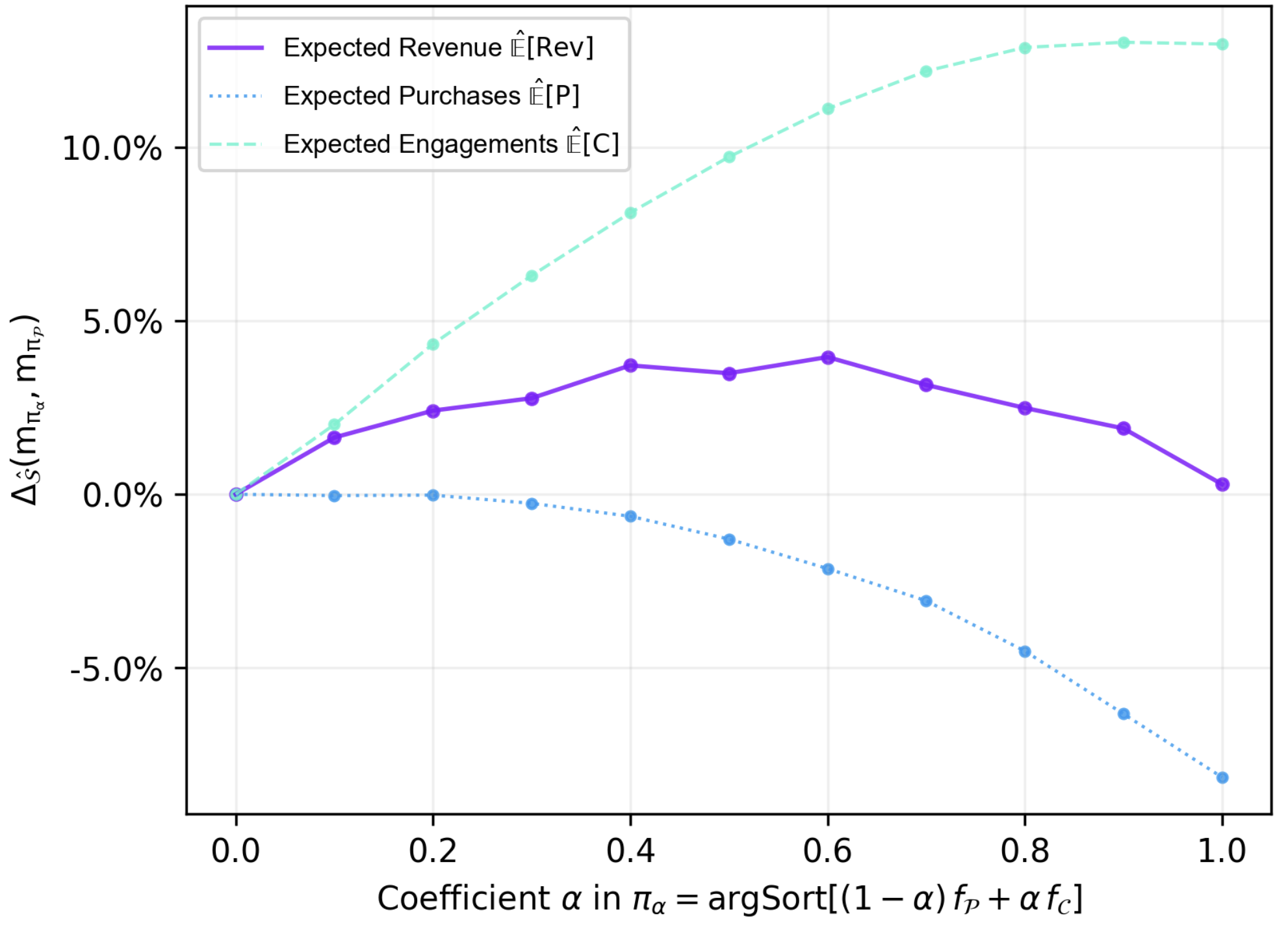}\vspace{-0.25cm}
    \caption{\small Counterfactual expected reward estimates $\Delta_{\hat{\mathcal{S}}}(m_{\pi_{\mathcal{\alpha}}},m_{\pi_{\mathcal{P}}})$ as a function of parameter $\alpha$ in .}
    \label{fig:total_reward_not_segmented}
\end{figure}
\vspace{-0.35cm}Figure \ref{fig:total_reward_not_segmented} highlights the essential trade offs between different expected reward estimates $m$ from logged data $\hat{\mathcal{S}}$, with the acquisition focused policy as the baseline 
\begin{eqnarray}
\Delta_{\hat{\mathcal{S}}}(m_{\pi_{\mathcal{\alpha}}},m_{\pi_{\mathcal{P}}}).
\end{eqnarray}
Biasing the training objective heavily on one extreme, leads to significant drops in the estimated reward corresponding to the other extreme.
As the contribution of the engagement-focused policy increases, by increasing $\alpha>0$, we estimate higher expected number of engagements, with a saturation point of diminishing return, after which a sharp drop in the expected number of purchases is observed.
Interestingly the estimated expected revenue is convex as a function of $\alpha$, which we will discuss in our subsequent research focused on value-aware objectives.  

\begin{figure}[H]
    \includegraphics[scale=0.18]{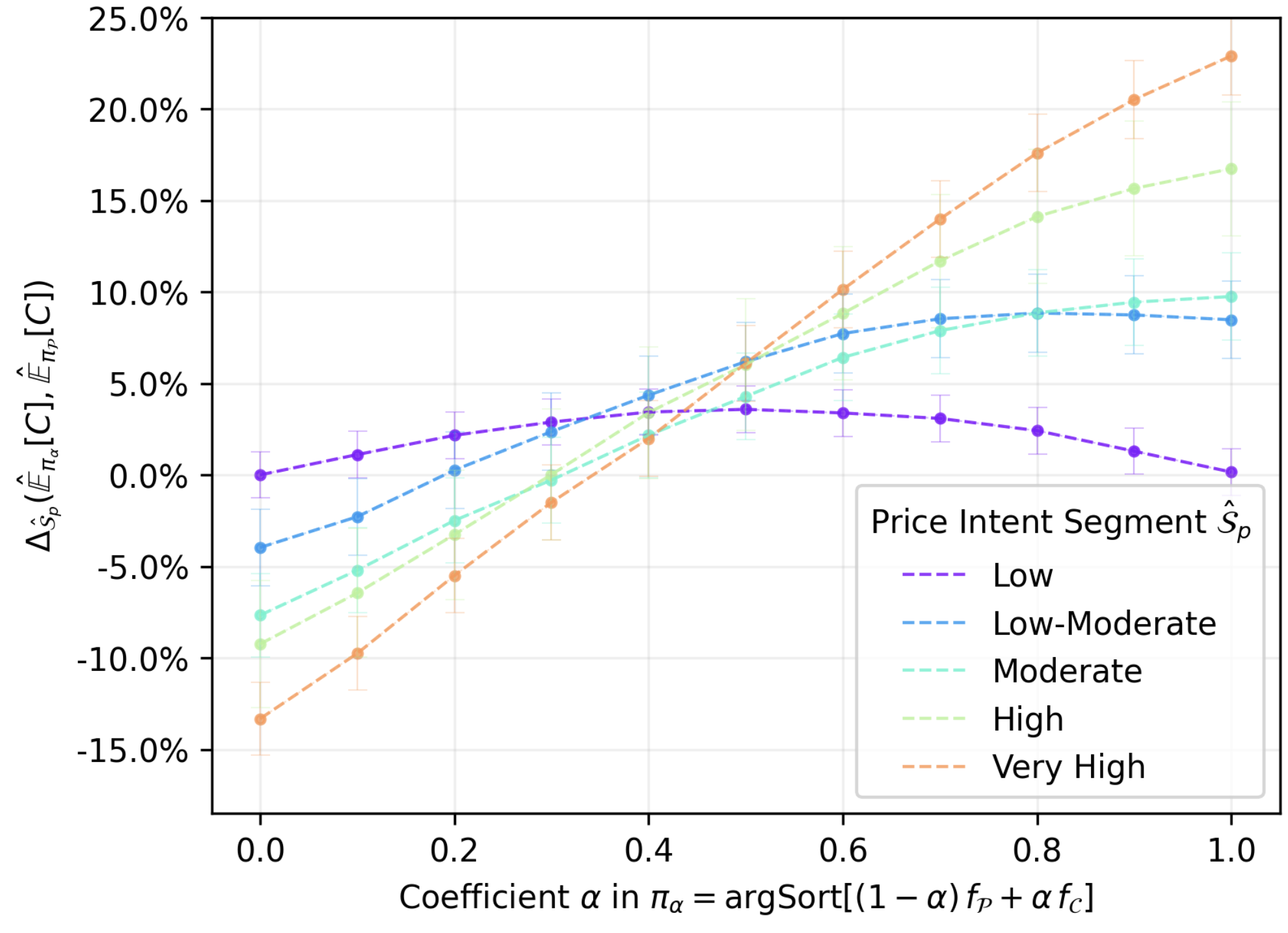}\vspace{-0.25cm}
    \caption{\small Counterfactual estimate for lift in the expected clicks $\Delta_{\hat{\mathcal{S}}_p}(\hat{\mathbb{E}}_{\pi_{\mathcal{\alpha}}}[\mathrm{C}],\hat{\mathbb{E}}_{\pi_{\mathcal{P}}}[\mathrm{C}])$ across price segments as a function of $\alpha$.}
    \label{fig:expengagement}
\end{figure}

\begin{figure}[H]
    \includegraphics[scale=0.18]{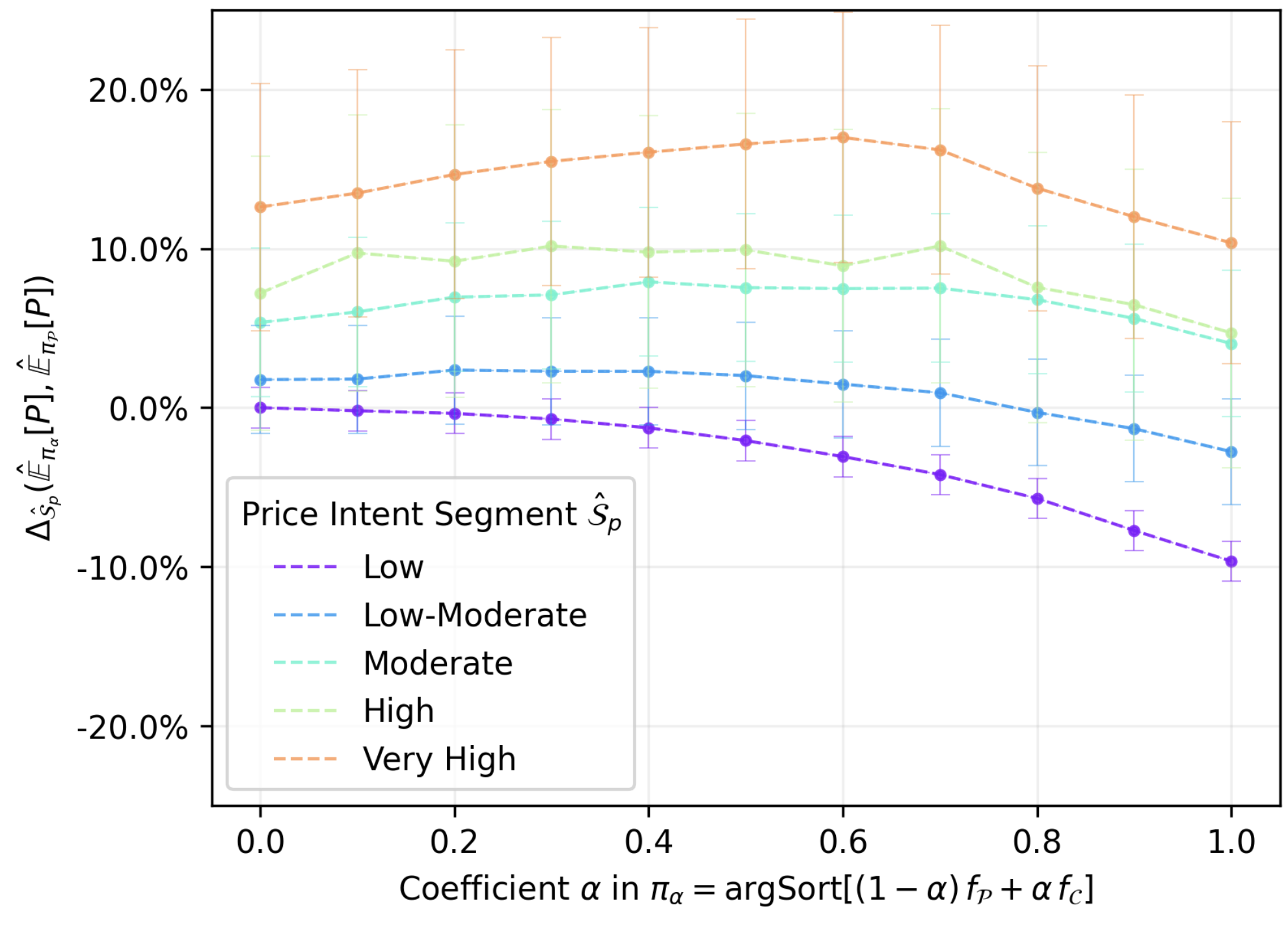}\vspace{-0.35cm}
    \caption{\small Counterfactual estimate for lift in the expected purchases $\Delta_{\hat{\mathcal{S}}_p}(\hat{\mathbb{E}}_{\pi_{\mathcal{\alpha}}}[\mathrm{P}],\hat{\mathbb{E}}_{\pi_{\mathcal{P}}}[\mathrm{P}])$ across price segments as a function of $\alpha$.}
    \label{fig:expsale}
\end{figure}

Next, we explore the observed trade-offs in the global analysis above across heterogeneous segments $\hat{\mathcal{S}}_p$ corresponding to different price intent segments, where the attribution of a session to a value bucket is done with respect to the price of the purchased items. Figure \ref{fig:expengagement} and \ref{fig:expsale} show the lift in estimated expected number of engagements $\hat{\mathbb{E}}[\mathrm{C}]$ and estimated expected number of engagements $\hat{\mathbb{E}}[\mathrm{P}]$, respectively, for the hybrid policy $\pi_{\mathcal{\alpha}}$ across value segments $\hat{\mathcal{S}}_p$ with the acquisition focused policy as the baseline. 

We clearly see that the extreme acquisition focused policy performs poorly in terms of the expected number of engagements, across all segments, with particularly larger effects sizes in high value price intents that require more exploration.
We also observe that, as $\alpha$ increases, the lift in expected clicks $\Delta_{\hat{\mathcal{S}}_p}(\hat{\mathbb{E}}_{\pi_{\mathcal{\alpha}}}[\mathrm{C}],\hat{\mathbb{E}}_{\pi_{\mathcal{P}}}[\mathrm{C}])$ increases, with a saturation point in lower price segments(which is in fact an inflation point for low price intent segments). 
On the contrary, biasing the policy towards the engagement-focused policy, by setting $\alpha$ close to 1, leads to a meaningful drop in the expected number of purchases, $\Delta_{\hat{\mathcal{S}}_p}(\hat{\mathbb{E}}_{\pi_{\mathcal{\alpha}}}[\mathrm{P}],\hat{\mathbb{E}}_{\pi_{\mathcal{P}}}[\mathrm{P}])$, particularly in low value price segments, which constitute a high proportion of the overall number of purchases.
An interesting observation, however, is that focusing more on an engagement-based objective is helpful for driving even higher expected number of purchases in higher price segments.
We leave deeper dives on the observed trade offs for future work.\vspace{-0.15cm}
\vspace{-0.35cm}
\subsubsection{Tight Attribution of Purchase Events}
In order to highlight the significance of the reward attribution scheme within a user session, we contrast the generalization performance of policies trained with respect to extreme choices of the query contribution distribution $\hat{\mathbb{P}}_{\mathcal{P}}(q|s_{\prec q})$.
Specifically, we contrast the performance of a policy $\pi_{t}$ trained on a session value distribution with a tight attribution of success events to search events, similar to the last touch scheme discussed earlier, to a policy $\pi_{l}$ trained with respect to a loose multi-touch attribution of success events to search events, similar to the all touch scheme discussed earlier.\vspace{-0.25cm}   
\begin{table}[H]
    \centering
    \begin{tabular}{| c  | c | c | c |}
     \multicolumn{1}{c}{} & \multicolumn{3}{c}{$\Delta_{\mathcal{S}_p}(m_{\pi_{l}},m_{\pi_{t}})$} \\ \hline
      Price Intent Buckets $p$  & Purchases & Engagements & Revenue  \\ \hline
      Low  &  -0.95\% & -0.66\% & -0.68\% \\ \hline
      Low-Moderate  &  +0.61\% & -0.13\% & -0.14\% \\ \hline
      Moderate   & +0.53\% & +0.44\%  & -0.41\%\\ \hline
      High  & +4.10\% & +0.94\% & +1.35\%\\ \hline 
      Very High  & +3.89\% & +0.74\% & +2.11\% \\ \hline
    \end{tabular}
    \caption{\small Online AB Test Results contrasting policies trained on extreme success attribution schemes}
    \label{tab:AB_attribute_PriceBucket}
\end{table}\vspace{-0.75cm}
While the overall cumulative rewards do not show sizable performance trade-offs between the two extreme policies, we highlight substantially different effect sizes across different purchase price intents. Table \ref{tab:AB_attribute_PriceBucket} clearly demonstrates that a loose attribution scheme for the empirical query contribution distribution helps with a significant improved generalization in higher price intent sessions, which tend to be more exploratory and involve multiple ranking intervention touch points.
\subsubsection{Purchase Price in Marketplace Reward}
Finally, in order to highlight the significance of incorporating the purchase price in the session value distribution for a revenue focused marketplace reward, we highlight the results from an online AB test on a simple value-aware policy $\pi_{\text{v}}$ in contrast to a value oblivious acquisition driven policy $\pi_{\text{\sout{v}}}$.
The primary difference between the two policies is the empirical session value distribution $\hat{\mathbb{P}}(s)$ in the corresponding expected reward estimate for the training objective, which depends also on the price of the sold item in the case of $\pi_{\text{v}}$, and all the other importance weighting distributions and per query reward estimates are the same.\vspace{-0.25cm} 
\begin{table}[H]
    \centering
    \begin{tabular}{| c  | c |}
    \hline
     Price Intent Buckets $p$  & $\Delta_{\mathcal{S}_p}(\mathrm{Rev}_{\pi_{\text{v}}},\mathrm{Rev}_{\pi_{\text{\sout{v}}}}$) \\ \hline
      Low  &  -1\% \\ \hline
      Low-Moderate  &  -0.1\%\\ \hline
      Moderate   &  +0.5\% \\ \hline
      High  & +1.3\% \\ \hline 
      Very High  & +1.9\% \\ \hline
    \end{tabular}
    \caption{\small Online AB test results contrasting a purchase value-aware policy against a value oblivious policy}
    \label{tab:AB_PriceWeight_PriceBucket}
\end{table}
\vspace{-0.75cm}
In Table \ref{tab:AB_PriceWeight_PriceBucket}, we clearly see a significant shift in the distribution of the accumulated revenue across price intent segments, which signifies the importance of taking into account the sparsity of purchase events from higher price intent sessions to avoid the selection bias incurred by the value oblivious policy, which is biased towards lower price intent segments, where sessions with purchase events are abundant.

\section{Concluding Remarks}
We established an explicit connection between the training objective for the search ranking policy and the key performance metrics of a two-sided commerce marketplace by building effective empirical estimates of the marketplace reward from observation data. 
Specifically, we highlighted the significance of the \textit{search context value distribution} in building effective empirical estimates of the marketplace expected reward to inform the training and evaluation of the search ranking policy.
We showcased empirical results from online randomized controlled experiments and counterfactual evaluations in a major e-commerce platform demonstrating the fundamental trade-offs governed by extreme choices of the context value distribution.
\bibliographystyle{ACM-Reference-Format}
\bibliography{PolicyExpectedValue}

\end{document}